\documentclass[12pt]{iopart}
\usepackage{graphicx}
\usepackage{epstopdf}
\usepackage{color}
\usepackage{multicol}

\usepackage{bm}
\usepackage[final]{pdfpages}

\newcommand{\myfigure}{./}
\newcommand{\revise}{\color{black}}

\begin{document}
	\title{Effects of anisotropic energetic particles on zonal flow residual level
	}
	\author{Z. X. Lu$^{1}$, M. Weiland$^{1}$, Ph. Lauber$^{1}$, X. Wang$^{1}$, G. Meng$^{1}$, F. Zonca$^{2}$}
	\address{$^1$ Max Planck Institut f\"ur Plasmaphysik, Garching, Germany \\
		$^2$ ENEA, Fusion and Nuclear Safety Department, C. R. Frascati, Via E. Fermi 45, 00044 Frascati (Roma), Italy
	}
	\date{\today}
	\begin{abstract}
		In tokamak plasmas, the interaction among the micro-turbulence, zonal flows (ZFs) and energetic particles (EPs) can affect the turbulence saturation level and the consequent confinement quality and thus, is important for future burning plasmas. In this work, the EP anisotropy effects on the ZF residual level are studied by using anisotropic EP distributions with dependence on pitch. Significant effects on the long wavelength ZFs have been found when small to moderate width around the dominant pitch in the EP distribution function is assumed. In addition, it is found that ZF residual level is enhanced by barely passing/trapped and/or deeply trapped EPs, but it is suppressed by well passing and/or intermediate trapped EPs. Numerical calculation shows that for ASDEX Upgrade plasmas, typical EP distribution functions can bring in $-3\%\sim+5.5\%$ mitigation/enhancement in ZF residual level, depending on the EP distribution functions.
	\end{abstract}
	
	
	\section{Introduction}
\begin{figure}\centering
		\includegraphics[width=0.48\textwidth]{\myfigure/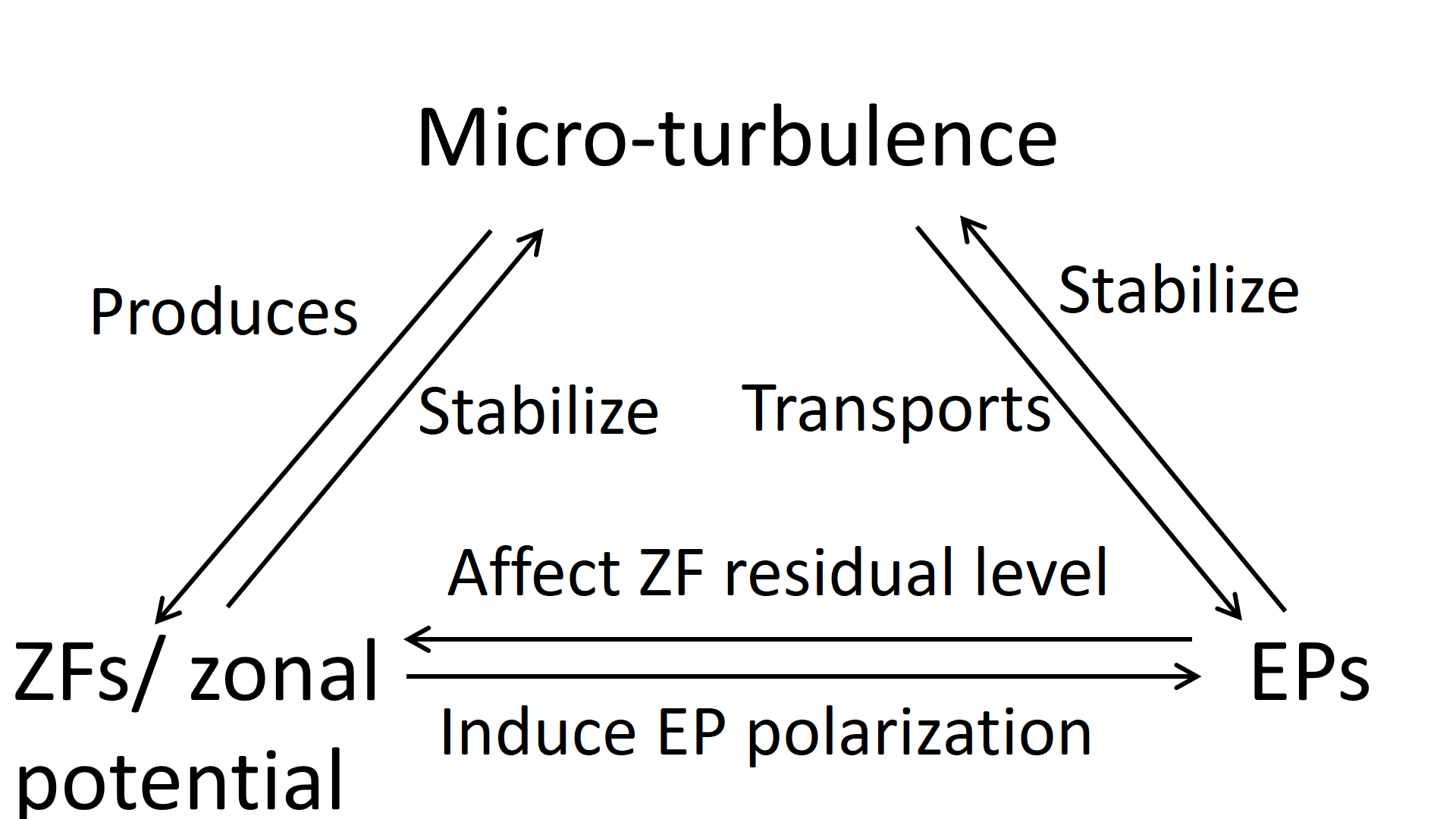} 
		\caption{Interplay of micro-turbulence, ZFs and EPs. }\label{fig:zf_ep_turb}
	\end{figure}

	In tokamak plasmas, zonal flows (ZFs) can regulate the micro-turbulence and reduce the transport. In burning plasmas or the present experimental plasmas with fast particles, the interplay among micro-turbulence, ZFs and energetic particles (EPs) can affect the eventual confinement. 
	As shown in Fig. \ref{fig:zf_ep_turb}, the interaction between ZFs and micro-turbulence has been intensively studied theoretically  \cite{diamond2005zonal,chen2000excitation,chen2007nonlinear} and numerically \cite{lin1998turbulent}. 
	The effect of micro-turbulence on EP transport has been demonstrated in gyrokinetic simulations \cite{zhang2008transport}. 
	In turn, EPs' effect on micro-turbulence suppression has been suggested by gyrokinetic simulations using realistic EP distribution functions \cite{di2018non}. 
	{\revise Recent simulations using ASDEX Upgrade H-mode parameters suggest that a high-confinement regime can be generated in the core with fast particles due to a full suppression of the turbulence by strong shear  flows \cite{di2020new}. However, the EP effects on ZF residual has not been discussed in this previous work. 
	}  
	One basic element of the EP-ZF interaction is related to the ZF residual level. 
	{\revise
		In axisymmetric  plasmas, ZFs are not completely damped by collisionless processes but stay in a finite level, namely, the ZF residual level \cite{rosenbluth1998poloidal}. This ZF residual level ($R_{ZF}$) is determined by the classical polarization density ($n^{pol}_{cl}$) due to the finite Larmor radius effect and the neoclassical polarization density ($n^{pol}_{nc}$) due to the finite orbit width effect, namely, $R_{ZF}=n^{pol}_{cl}/(n^{pol}_{cl}+n^{pol}_{nc})$. }After studies of the thermal ions' effect on ZF residual level \cite{rosenbluth1998poloidal,xiao2006short}, the impurity ions' effect on ZF has been also studied for multiple species of Maxwellian distribution with different masses and temperatures \cite{hahm2013isotopic,guo2017impurity} based on the generalized expression for the polarization density \cite{wang2009generalized}, where EPs are viewed as an isotropic impurity. Besides the isotope effects \cite{hahm2013isotopic}, the effect of the temperature anisotropy on ZF residual level \cite{cho2021effect}, the ZF residual level in stellarators \cite{monreal2016residual} and  the electromagnetic effects on ZFs \cite{catto2017electromagnetic} have been also investigated. In our previous work, the EP anisotropic effects have been studied based on analytical model EP distribution functions \cite{lu2019theoretical}.
	
	In this work, the effects of anisotropic EPs on ZF residual level are studied and applied to the analyses using {\revise realistic EP distributions in ASDEX Upgrade discharges and the ZF residual level solver developed recently}. The motivation and the scope of this work include,
	\begin{enumerate}
		\item the derivation of the ZF residual level for arbitrary distribution functions with multiple species included and the numerical implementation, as a complement of the previous theoretical and numerical work using the Maxwellian distribution function \cite{hahm2013isotopic,xiao2006short} or the bi-Maxwellian distribution function \cite{cho2021effect};
		\item the exploration of the theoretical foundation of the ZF control using anisotropic particles induced by the neutral beam injection (NBI) or ion cyclotron resonance heating (ICRH);
		\item the identification of the effect of anisotropic EPs on ZF residual level by making use of 
		more realistic analytical and numerical treatment of EP distribution functions, following our previous work \cite{lu2019theoretical};
		\item the prediction of anisotropic EP effects in burning plasma or high $\beta$ plasma, in addition to the present studies related to  isotope effects \cite{hahm2013isotopic,garcia2018isotope}.
	\end{enumerate}
	
	This work is organized as follows. In Section \ref{sec:equations}, the equation for the calculation of ZF residual level is {\revise demonstrated} with arbitrary distribution functions taken into account. 
	{\revise
		In Section \ref{sec:short_wavelength}, the ZF residual level for arbitrary wavelength is calculated numerically and the passing/trapped EPs' effects on ZF residual level are demonstrated, using the model EP distribution and realistic EP distributions from ASDEX Upgrade discharges. 
	}
	In Section \ref{sec:conclusion}, the summary and the outlook are discussed.
	
	\section{Zonal flow residual level for arbitrary distribution functions}\label{sec:equations}
	\subsection{General description of the zonal flow residual level}
	The linear response of plasmas to electrostatic perturbation can be obtained by solving the linearized gyrokinetic equation with the separation between the adiabatic and non-adiabatic responses,
	\begin{eqnarray}
	\delta f_s = -\frac{e_s\delta\phi}{T_{0s}}\kappa F_{0s}+\delta K_s e^{iL_{ks}}\;\;, 
	\end{eqnarray}
	where $s$ is the species index, $e_s$ the electric charge of species $s$, {\revise $T_{0s}$ the temperature, $\delta\phi$ the perturbed scalar potential, $\delta K_s$ the non-adiabatic part of the perturbed distribution function, } $F_{0s}$ the equilibrium distribution function, 
	{\revise $\kappa$ is defined according to $\kappa F_{0s}\equiv -(T_{0s}/m_s)\partial F_{0s}({\bf R},E,\mu)/\partial{E}$,  $m_s$ the mass, }
	$\mu=v_\perp^2/(2B)$,  ${E}=v^2/2$, $L_{ks}={\bf k}\times{\bf b}\cdot{\bf v}/\omega_{cs}$, $\omega_{cs}=e_sB/(cm_s)$, ${\bf b}={\bf B}/B$, $\bf B$ is the equilibrium magnetic field. {\revise For Maxwellian distribution function, $\kappa=1$ from the definition of $\kappa F_{0s}$. } Note that for an arbitrary distribution function, $T_{0s}$ is used as the reference quantity instead of the temperature of species $s$. All perturbations are written in the eikonal form for the description of the perpendicular (to $\bf b$) variation as a specific application of the Mode Structure Decomposition (MSD) approach \cite{zonca1992resonant,lu2012theoretical,lu2013mixed}, e.g., $\delta\phi({\bf r})=\delta\phi_k\exp\{iS({\bf x_\perp})\}$.
	The gyrokinetic and Poisson equations are adopted as a simplified version of the electromagnetic gyrokinetic equations \cite{antonsen1980kinetic,chen1991kinetic,zonca1996kinetic},
	\begin{eqnarray}
	\label{eq:gk0}
	&&\left[\partial_t+v_{||}\partial_{||}+i\omega_{d,s}\right] \delta K_s 
	= \frac{e_s}{T_{0s}} \kappa F_{0s}  J_0(a_s)\partial_t\delta\phi\;\;, \\
	\label{eq:qn0}
	&&\left[ \sum_{s=e,i,f}\left(\int dv^3\kappa\bar{F}_{0s}\right)\frac{n_{0s} e_s^2}{T_{0s}}  \right] \delta\phi
	= \sum_{s=e,i,f}\left(e_s\int dv^3  J_0(a_s) \delta K_s \right)\;\;,
	\end{eqnarray}
	where the independent velocity variables ${E}$ and $\mu=v_\perp^2/(2B)$ are used, $\bar{F}_{0s}=F_{0s}/n_{0s}$, $\partial_{\parallel}={\bf b}\cdot\nabla$, 
	$\omega_{d,s}=-i{\bf{v}}_{d,s}\cdot \nabla$,  ${\bf{v}}_{d,s}=-v_\parallel{\bf{b}}\times\nabla_{{E},\mu}(v_\parallel/\omega_{cs})$,  $a_s=k_\perp\rho_s$ (for ZFs, $k_\perp=k_r$), $\rho_s=v_\perp/\omega_{cs}$, equilibrium density and temperature profiles are assumed to be uniform in radial direction. 
	The source term in Eq. \ref{eq:gk0} is not explicitly written compared with \cite{chen2000excitation,rosenbluth1998poloidal} since we focus on the calculation of the polarization/neoclassical density for an arbitrary distribution function that gives the ZF residual level.
	In the following, the subscript `s' is omitted when no ambiguity is introduced. 
	Equation \ref{eq:gk0} is solved in the low frequency limit ($\omega/\omega_{tr}\ll1$, where $\omega_{tr}=v_\parallel/(qR)$, $\omega$ is the mode frequency), as shown in \ref{app:gkf1derivation}.  
	{\revise The integral of $\delta f$ in velocity space gives the total polarization density, }
	\begin{eqnarray}\label{eq:fpoltot0}\fl\qquad
	\left\langle\delta n^{pol}\right\rangle 
	=  \frac{en_0\delta\phi}{T_{0}} \left\langle
	\int dv^3\left(
	J_0e^{-iQ} \overline{J_0 e^{iQ}}
	-1\right)\kappa\bar F_0
	\right\rangle\;\;,
	\end{eqnarray}
	{\revise
		where $Q$ is due to the finite orbit width effect and is defined in \ref{app:compare_RH},
	}
	$\langle\ldots\rangle$ indicates the flux surface average and the bounce average operation is defined as $\bar{A}=\oint (dl/v_\parallel)A/\oint (dl/v_\parallel)$. The classical polarization density is 
	\begin{eqnarray}\label{eq:fpolcl0}\fl\qquad
	\left\langle\delta n^{pol}_{cl}\right\rangle
	=\frac{en_0\delta\phi}{T_{0}} \left\langle
	\int dv^3\left(
	J_0^2
	-1\right)\kappa \bar F_0
	\right\rangle\;\;.
	\end{eqnarray}
	The neoclassical polarization can be also obtained $\langle\delta n_{nc}^{pol}\rangle=\langle\delta n_{}^{pol}\rangle-\langle\delta n_{cl}^{pol}\rangle$.
	Equations \ref{eq:fpoltot0} and \ref{eq:fpolcl0} can be reduced to the previous results for Maxwellian distribution ($\kappa F_0=F_0$) \cite{rosenbluth1998poloidal,xiao2006short}. 
	{\revise 
		The ZF residual level for multiple species is \cite{hahm2013isotopic,guo2017impurity}
		\begin{eqnarray}\label{eq:ResidualZF0}
		R_{ZF} 
		= \frac{\sum_s e_s\left\langle \delta  n^{pol}_{cl} \right\rangle}{\sum_s e_s\left\langle \delta  n^{pol} \right\rangle}\;\;.
		\end{eqnarray}
	}
	\subsection{EPs' effect on ZF residual level in the long wavelength limit}
	\label{sec:long_wavelength}
	{\revise
		The ZF residual level is derived analytically in the long wave length limit using the single pitch EP distribution, for identifying the underlying physics related to the negative polarization. }
	Equations \ref{eq:fpoltot0} and \ref{eq:fpolcl0} can be solved in the long wave length ($k_r\rho_s\ll1$)  limit, for which 
	{\revise
		$J_0(a)\approx1-a^2/4$, $\exp\{\pm iQ\}\approx1\pm iQ-Q^2/2$. 	
	} 
	Noticing that
	$\int dv^3=({\sqrt{2}\pi}/{h})\sum_{\sigma}\int_{0}^\infty dE\int_{0}^{h}d\lambda{\sqrt{E}}/{\xi}$, $\xi\equiv {|v_\parallel|}/{v}=\sqrt{1-\lambda/h}$,
	$\lambda=v^2_\perp B_0/(v^2B)$ is the pitch,  
	$\sigma$ is the sign of $v_\parallel$, $h=1+\varepsilon\cos\theta$, $\varepsilon=r/R_0$, Eqs. \ref{eq:fpoltot0}--\ref{eq:fpolcl0} yields  
	\begin{eqnarray}
	\label{eq:npol0}\fl\quad
	\left\langle n^{pol}_{nc} 	\right\rangle=
	-n_0\frac{e\delta\phi}{T_0} 
	\left(\frac{qk_r}{\varepsilon\omega_{c0}}\right)^2 
	2\sqrt{2}\pi\sum_{\sigma=\pm1}\int d\lambda
	\oint \frac{d\theta}{2\pi}\int_{0}^{\infty}dE\kappa\bar{F}E^{3/2}
	\left[
	h^2\xi-\frac{2\pi h}{\oint\frac{d\theta}{\xi}}
	\right]\;\;,\\
	\label{eq:nclpol0}
	\fl\quad
	\left\langle n^{pol}_{cl} 	\right\rangle=-n_0\frac{e\delta\phi}{T_0} 
	\left(\frac{k_r}{\omega_{c0}}\right)^2 
	\sqrt{2}\pi\sum_{\sigma=\pm1}
	\oint \frac{d\theta h}{2\pi}\int_{0}^{\infty}dE\kappa\bar{F}E^{3/2}
	\left(
	\frac{1}{\xi}
	-\xi
	\right)
	\;\;.
	\end{eqnarray}
	{\revise
		Equations \ref{eq:npol0} and \ref{eq:nclpol0} are consistent with that for Maxwellian distribution ($\kappa=1$) \cite{rosenbluth1998poloidal,hahm2013isotopic,guo2017impurity,xiao2006short,xiao2007effects,garbet2016relationship}, as shown in \ref{app:compare_RH}. 
	}
	
	To identify the EPs' effects on ZF residual level in an explicit way, we choose the Maxwellian distribution in $v$ direction but a single pitch $\lambda_0$ for EPs, 
	{\revise
		\begin{eqnarray}
		\label{eq:FMP0}
		F_{MP}=C_P\delta(\lambda-\lambda_0) e^{-2\bar{E}}\;\;,\;\;
		C_P=\frac{2T_1(\lambda=\lambda_0)n}{\pi^{3/2}v_{T}^3}\;\;,
		\end{eqnarray}
		where $\bar{E}=E/v_{T}^2$ and $n$ is the flux surface-averaged density.
		From Eq. \ref{eq:dfdE2dfdlambda}, we have
		\begin{eqnarray}
		\kappa F_{MP}=F_{MP}+(1/2)(\lambda/\bar{E})[\partial_\lambda F_\lambda(\lambda)]C_P\exp[-2\bar{E}]\;\;.
		\end{eqnarray}
	}
	In addition, we choose isotropic Maxwellian thermal ions. Equation \ref{eq:ResidualZF0} yields,
	\begin{eqnarray}\label{eq:ResidualZF1}
	R_{ZF}=
	\frac{n_i\left(1+\frac{3}{2}\varepsilon^2\right)+n_fP_{cl}}
	{n_im\left(1+\frac{3}{2}\varepsilon^2\right)+n_fP_{cl}
		+\frac{q^2}{\varepsilon^2}(n_iI_{tot}+n_fP_{nc})}\;\;, \\
	\label{eq:varepscl_PM0}
	P_{cl}(\lambda_0)=-\left[T_1\left(\frac{1}{2}-\lambda\partial_{\lambda}\right)\left(T_2-\frac{1}{T_1}\right)\right]_{\lambda_0}\;\;,
	\\
	\label{eq:vareps_PM0}
	P_{nc}(\lambda_0)=\left\{
	T_1\left[
	T_2 -2\lambda\partial_\lambda T_2
	-\sigma_p(T_1 -2\lambda\partial_\lambda T_1)
	\right]\right\}_{\lambda=\lambda_0}\;\;.
	\end{eqnarray}
	where 
	{\revise it is assumed that EPs and thermal ions are the same species, }
	the subscript `$f$' indicates EPs (fast particles), $T_1$, $T_2$ and $I$ are defined in Eqs. \ref{eq:vareps0}--\ref{eq:T12},
	{\revise 
		and $P_{cl/nc}=\varepsilon^{pol}_{cl/nc}/n_{s0}$, the polarization $\varepsilon^{pol}_{cl/nc}$ is defined in Eq. \ref{eq:define_polarization}. For Maxwellian distribution, the classical and neoclassical polarization are positive. For the anisotropic distribution function, $\partial T_1/\partial\lambda$ and  $\partial T_2/\partial\lambda$ in Eqs. \ref{eq:varepscl_PM0}--\ref{eq:vareps_PM0} origin from $\partial f_0/\partial \lambda$ in the gyrokinetic equation and  introduce additional terms which can lead to negative polarization. Generally, the sign of the polarization can be positive or negative, depending on the particle distribution function $F_0$. Strong sources can lead to the deviation of $F_0$ away from Maxwellian and can introduce the inverted gradients in phase space and thus can alter the sign of the polarization (``negative polarization shielding''). Note in this work, only $n=0$ zonal component is included and other instabilities such as Alfv\'enic modes that can be driven by the non-Maxwellian distribution are not considered. 
	}
	
\begin{figure}\centering
		\includegraphics[width=0.98\textwidth]{\myfigure/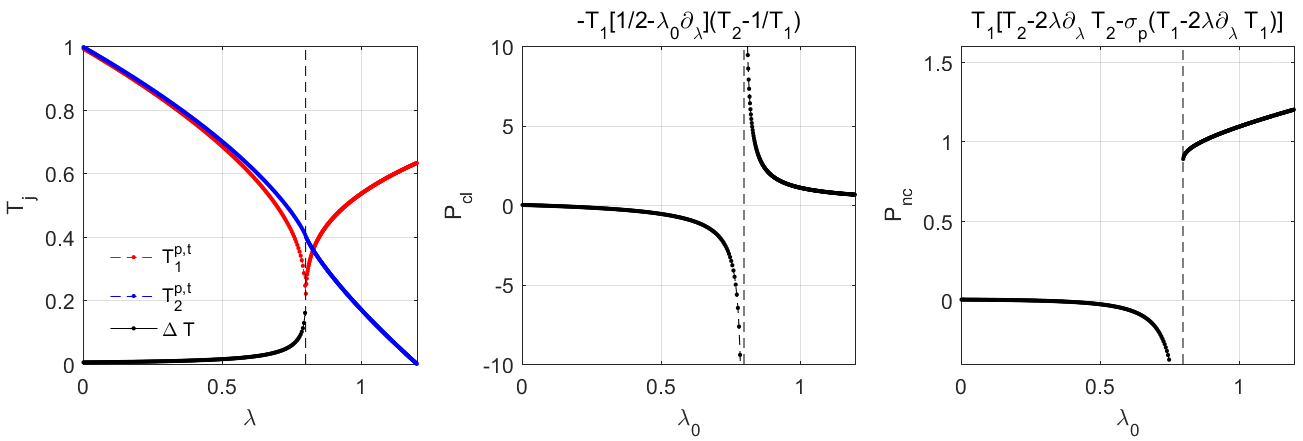} 
		\caption{The response function for $\varepsilon=0.2$. Left: the integral kernel defined in Eq. \ref{eq:T12}; central and right: the coefficients in classical and neoclassical polarization for Maxwellian distribution with a single pitch $\lambda_0$ defined in Eqs. \ref{eq:varepscl_PM0} and \ref{eq:vareps_PM0}. {\revise The vertical dashed line indicates the passing-trapped boundary. } }\label{fig:TSplot}
	\end{figure}

	The plasma density response to zonal scalar potential $\delta\phi$ is described by $(T_1,T_2)$ in Eq. \ref{eq:T12} and $(P_{cl}, P_{nc})$ in Eqs. \ref{eq:varepscl_PM0}--\ref{eq:vareps_PM0}, and determines the ZF residual level according to Eq. \ref{eq:ResidualZF1}. Figure \ref{fig:TSplot} shows the dependence of $T_1$, $T_2$, $P_{cl}$ and $P_{nc}$ on $\lambda_0$. 
	{\revise In the left frame,} the results of $T_{1}^p$, $T_2^{p,t}$ are consistent with those in \cite{xiao2007collisional}. 
	For classical and neoclassical polarization shown in the central and right frames, the curve is discontinuous at the passing-trapped boundary, indicating the different behaviors of passing and trapped particles. As $\lambda_0$ increases, the value of $P_{cl}$ decreases for either passing or trapped particles. 
	While for a single particle, the classical polarization increases in the whole $\lambda$ range since $1-J_0^2(k_\perp\rho)\approx  h k_\perp^2E\lambda/\omega_{c0}^2$, the dependence of $P_{cl}$ on $\lambda_0$ is also contributed by the phase space gradient (the $\partial_\lambda$ term) and the normalization factor $C_P$. 
	In the right frame, the negative and positive values of $P_{nc}$ for passing and trapped particles indicate  positive and negative contributions to $R_{ZF}$ respectively according to Eq. \ref{eq:ResidualZF1}, given that the contribution from EPs is perturbative ($n_f/n_i\ll1$). {\revise The negative and positive $P_{cl}$ values indicate negative and positive contributions to $R_{ZF}$ and thus the net effects from EPs on $R_{ZF}$ rely on the competition between $P_{cl}$ and $P_{nc}$. }
	
\begin{figure}\centering
		\includegraphics[width=0.48\textwidth]{\myfigure/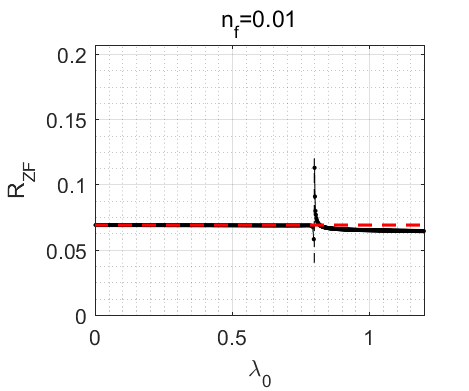}
		\includegraphics[width=0.48\textwidth]{\myfigure/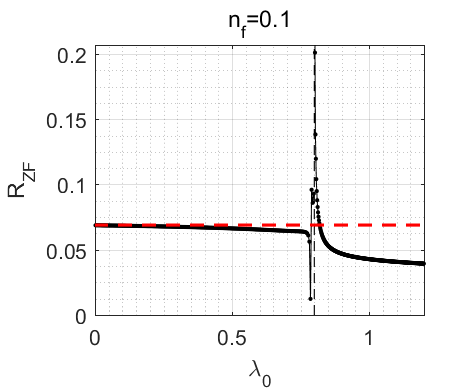} 
		\caption{ZF residual for $n_f/n_e=0.01$ (left) and $n_f/n_e=0.1$ (right).}\label{fig:ZF_residual}
	\end{figure}

	The ZF residual level $R_{ZF}$ is calculated and shown in Fig. \ref{fig:ZF_residual}. With passing EPs, $R_{ZF}$ decreases as $\lambda_0$ increases and $R_{ZF}$ is smaller than the Rosenbluth-Hinton (R-H) result for pure thermal ions (red dashed line), unless EPs are close to the passing-trapped boundary where $R_{ZF}$ changes its sign. The barely trapped EPs enhance the ZF residual level while in the other range, the  trapped EPs can suppress $R_{ZF}$. As EP density increases from $n_f=0.01$ (left) to $n_f=0.1$ (right), the deviation away from the R-H result becomes more significant.
	
	\section{Numerical studies using model and experimental EP distributions}
	\label{sec:short_wavelength}
	{\revise
		\subsection{Equations and numerical implementation}
		\label{subsec:eq_numeric}
		For numerical implementation, the total polarization and the classical polarization in Eqs. \ref{eq:npol0}--\ref{eq:nclpol0} are written in $(E,\lambda)$ coordinates,
		\begin{eqnarray}\label{eq:pola_dens0}
		\fl\qquad
		\left\langle\delta{n}^{pol}\right\rangle &=& 
		\frac{e_sn_{0}\delta\phi}{T_0}\sqrt{2}\int_{0}^{\infty} dE \sqrt{E}\int d\lambda K({E},\lambda,Q)\kappa\bar{F}_0\;\;, \\
		\label{eq:KQ}
		\fl\qquad
		K(E,\lambda,Q)&=&2\oint \frac{d\theta}{\xi}(J_0\cos Q-1) \nonumber\\
		\fl\qquad
		&+&\left(\oint \frac{d\theta}{\xi}\right)^{-1}
		\left\{
		\left[\oint\frac{d\theta}{\xi}(J_0\cos Q-1)\right]^2
		+\left[\oint\frac{d\theta}{\xi}J_0\sin Q\right]^2
		\right\}\;\;,\\
		\label{eq:pola_cl_dens0}
		\fl\qquad
		\left\langle\delta n^{pol}_{cl}\right\rangle
		&=&\frac{en_0\delta\phi}{T_{0}} \sqrt{2}
		\int_{0}^{\infty} dE \sqrt{E}\int d\lambda
		\oint \frac{d\theta}{\xi} \left(J_0^2-1\right)\kappa \bar F_0\;\;,
		\end{eqnarray} 
		where $\exp\{\pm Q\}=\cos Q\pm i\sin Q$ has been used and the cancellation between $J_0\cos Q$ and $1$ has been considered \cite{xiao2006short}. 
	}
	In numerical implementation, Gaussian quadrature is used for the integral in $E$, $\lambda$ and $\theta$ directions in Eqs. \ref{eq:pola_dens0}--\ref{eq:pola_cl_dens0}. Integrals are calculated numerically using $\int_{x_a}^{x_b}f(x)dx\approx\sum_{i=1}^{n}c_if(x_i)$, where  $x_i$ and $c_i$ are nodes and weights using the Gaussian quadrature. 
	{\revise
		The integral along $\bar{E}$ is replaced with $\int d\bar{E}=\int \bar{v}d\bar{v}$, where $\bar{E}=\bar{v}^2/2$.
	}
	In order to eliminate the singularity in the denominator of the integral kernel, the integral along  $\theta$ is modified by defining the time-like variable $\tau$ according to,
	\begin{eqnarray}
	\tau_t=\sqrt{\frac{2}{\varepsilon}}F(\alpha,\kappa^2)\;\;, \;\;
	\tau_p=\frac{1}{\sqrt{2\varepsilon}}F(\theta,1/\kappa^2)\;\;,
	\end{eqnarray}
	for trapped particles and passing particles, respectively, where $\kappa\sin\alpha=\sin(\theta/2)$, $\kappa^2=(1+\varepsilon-\lambda)/(2\varepsilon)$, $F(a,b)$ is the incomplete elliptic integral of the first kind.
	
	\subsection{Parametric studies using model EP distribution functions}
	\label{subsec:model_EP}
\begin{figure}\centering
		\includegraphics[width=0.48\textwidth]{\myfigure/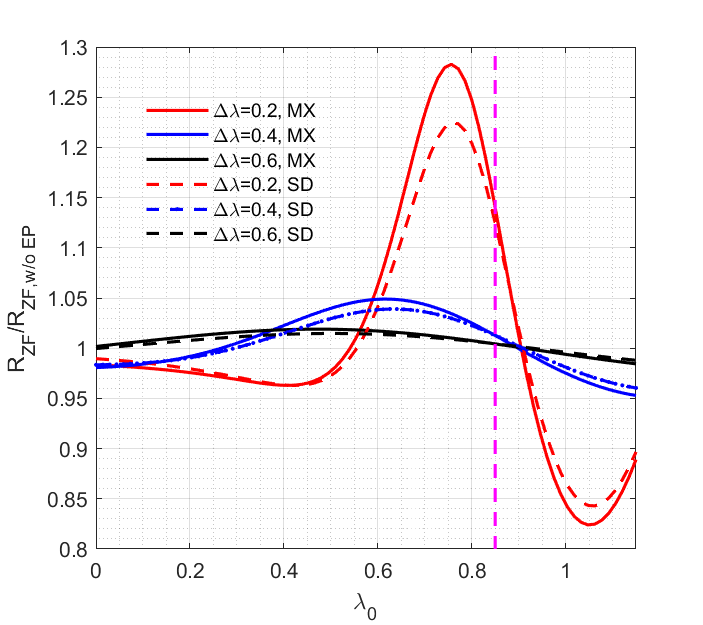}
		\caption{The ZF residual level for the model distributions with Maxwellian (MX) distribution along $v$ (solid lines) in Eq. \ref{eq:FMP1} and with slowing down (SD) distribution along $v$ (dashed lines) in Eq.  \ref{eq:FSD}. Other parameters are $\epsilon=0.15$, $q=2$, $k_r\rho_{Ti}=10^{-3}$, $n_f/n_e=0.1$, $T_f/T_i=10$.}\label{fig:max_slowdown_cmp}
	\end{figure}

	{\revise
		In order to demonstrate the dependence of the ZF residual level on EP anisotropy parameters, two types of model EP distributions are considered. The first one is the Maxwellian distribution in $v$ with finite width in pitch $\lambda$, 
		\begin{eqnarray}\label{eq:FMP1}
		\bar{F}_{MX}=C_{MX} \exp\{-2\bar{E}\}\exp\left\{-\left(\frac{\lambda-\lambda_0}{\Delta\lambda}\right)^2\right\}
		\;\;,\\
		C_{MX}=\left\{\sqrt{\pi}\int_{0}^{\pi}d\theta\int_{0}^1
		d\xi\exp\left[-\left[\frac{h(1-\xi^2)-\lambda_0}{\Delta\lambda}\right]^2\right]\right\}^{-1} \;\;,
		\end{eqnarray}
		where $C_{MX}$ (or $C_{DS}$ in Eq. \ref{eq:FSD}) is chosen so that $\oint d\theta h \int dv^3\bar{F}/\oint d\theta h=1$. 
		The other model distribution is the slowing down distribution in $v$ with finite width in pitch $\lambda$,
		\begin{eqnarray}\label{eq:FSD}
		\fl\qquad
		\bar{F}_{SD}=\frac{C_{SD}H(E_b-E)}{E^{3/2}+E_c^{3/2}} \exp\left\{-\left(\frac{\lambda-\lambda_0}{\Delta\lambda}\right)^2\right\}
		\;\;,\\
		\fl\qquad
		C_{SD}=\left\{\frac{8\sqrt{2}}{3}\ln\left[\left(\frac{E_b}{E_c}\right)^{3/2}+1\right]\int_{0}^{\pi}d\theta\int_{0}^1
		d\xi\exp\left[-\left[\frac{h(1-\xi^2)-\lambda_0}{\Delta\lambda}\right]^2\right]\right\}^{-1} \;\;,
		\end{eqnarray}
		where $E_b$ is the birth energy, the Heaviside function $H(x)=0$ for $x<0$ and $H(x)=1$ for $x>0$, $E_c$ is determined by the plasma parameters. 
		The ZF residual level for different values of $\lambda_0$ is calculated for $\Delta\lambda=0.2, 0.4, 0.6$, as shown in Fig. \ref{fig:max_slowdown_cmp}. We choose $k_r\rho_{Ti}=10^{-3}$ since the anisotropic effects are more significant in the moderate to long wavelength ($k_r\rho_{Ti}<0.1$) \cite{lu2019theoretical}. For Maxwellian EP distribution, $T_f/T_i=10$. For slowing down distribution $T_f/T_i=10$, $E_b=2T_f/m_f$, $E_c=0.2T_f/m_f$, $E_{b/c}=v_{b/c}^2/2$, where $v_b$ and $v_c$ are birth velocity and critical velocity respectively (note that the rigorous effective EP temperature is different than $T_f$ but varying $T_f$ in the long wavelength limit hardly changes ZF residual level). 
		For the slowing down EP distribution, the ZF residual level is closer to $1$ than that of the Maxwellian distribution and thus, the enhancement of the ZF residual level can be overestimated if Maxwellian distribution is adopted. However, the dependence on $\lambda_0,\Delta\lambda$ are similar for both model distributions. In the following, we choose the anisotropic Maxwellian distribution for parametric studies. More realistic calculation relies on the consideration of experimental EP distribution, as shown in Section \ref{subsec:experiment_EP}.
	}
	
\begin{figure}\centering
		\includegraphics[width=0.49\textwidth]{\myfigure/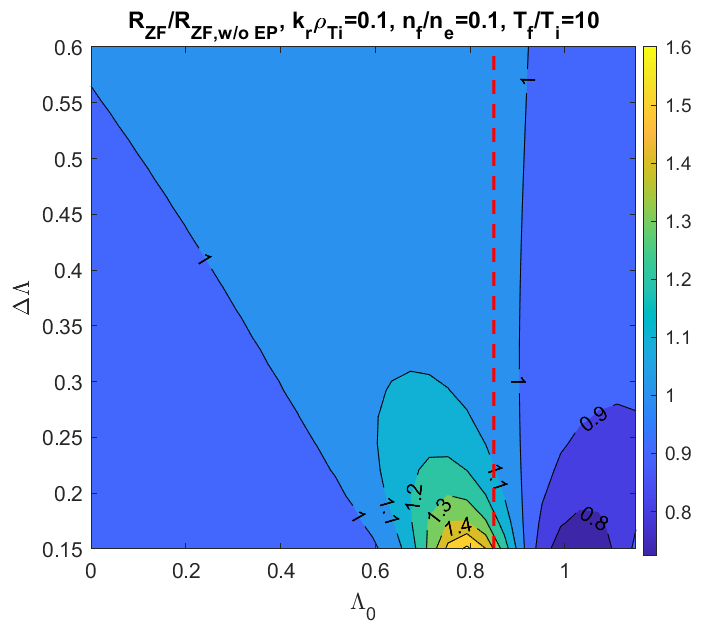}
		\includegraphics[width=0.49\textwidth]{\myfigure/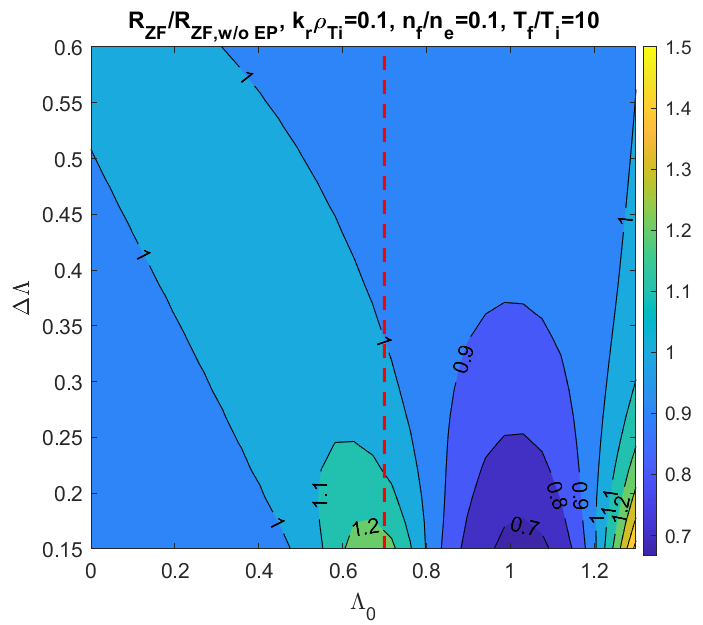}
		\caption{The ZF residual level $R_{ZF}$ versus EP parameters  $(\lambda_0,\Delta\lambda)$ for $\varepsilon=0.15$ (left) and $\varepsilon=0.3$ (right), given the EP distribution in Eq. \ref{eq:FMP1}.}\label{fig:ZF2d_lambda0_lambdawid}
	\end{figure}

\begin{figure}\centering
		\includegraphics[width=0.49\textwidth]{\myfigure/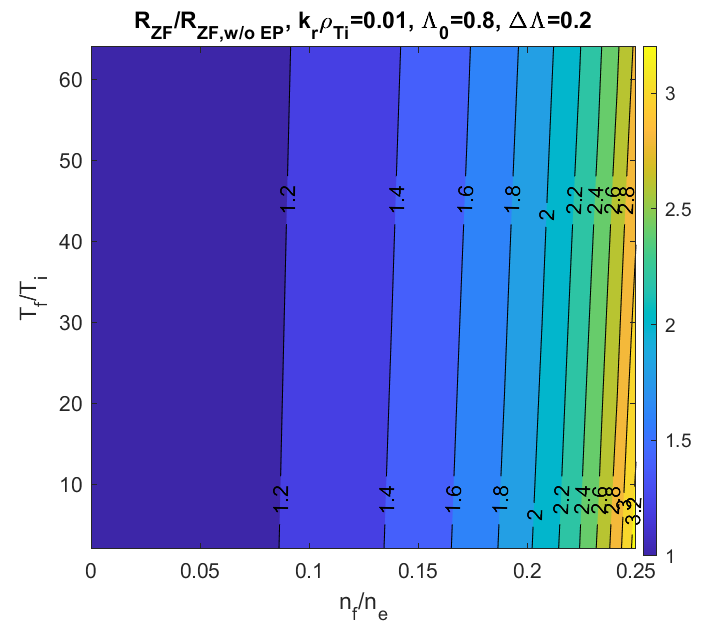}
		\includegraphics[width=0.49\textwidth]{\myfigure/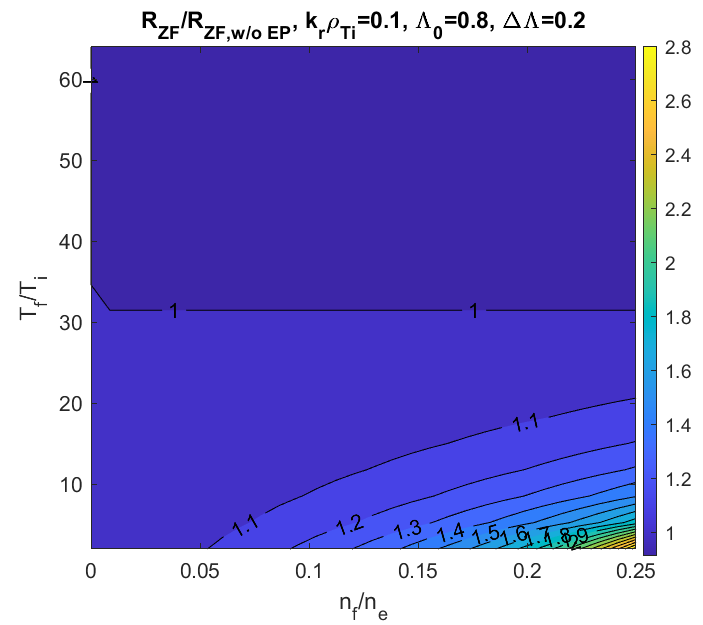}
		\caption{ZF residual level for different EP densities and temperature $T_f$ for $k_r\rho_{Ti}=0.01 (left)$ and $k_r\rho_{Ti}=0.1$ (right). Other parameters are $\lambda_0=0.8$, $\Delta\lambda=0.2$}\label{fig:cmp_nf_Tf}
	\end{figure}

	{\revise
		The overall EP anisotropy effects on ZF residual level can be demonstrated by the parametric studies in $(\lambda_0,\Delta\lambda)$ space for different values of $\varepsilon$. As observed in our previous work \cite{lu2019theoretical}, this effect is more significant for moderate to large scale ZF ($k_r\rho_{Ti}<0.1$ where $\rho_{Ti}=v_{Ti}/\omega_c$, $v_{Ti}=\sqrt{2T_i/m_i}$) and thus we take $k_r\rho_{Ti}=0.1$. The value of $\Delta\lambda$ describes the EP distribution width in $\lambda$ direction. As shown in Fig. \ref{fig:ZF2d_lambda0_lambdawid}, $R_{ZF}$ is affected by the EP parameter $\lambda_0$, considering several critical values, namely, the passing-trapped boundary $\lambda_c=1-\varepsilon$ and the maximum pitch $\lambda_m=1+\varepsilon$, where $\varepsilon=r/R_0$. The ZF residual level w/o EPs ($R_{ZF,w/o EP}$) is used as a baseline to evaluate the EP effects.  The left frame of Fig. \ref{fig:ZF2d_lambda0_lambdawid} shows $R_{ZF}/R_{ZF,w/o EP}$ for $\varepsilon=0.15$. $R_{ZF}/R_{ZF,w/o EP}$ is slightly suppressed compared with the R-H result when the applied EPs are dominated by passing particles except when $\lambda_0$ is very close to $\lambda_c$ ($0.6<\lambda_0<0.85$). For  trapped  EPs, $R_{ZF}$ is suppressed in the intermediate trapped region ($0.9<\lambda_0<1.1$). As $\lambda_0$ approaches $\lambda_0=1.15$, the ZF residual level increases due to the enhancement effects from deeply trapped particles. The right frame of Fig. \ref{fig:ZF2d_lambda0_lambdawid} shows $R_{ZF}/R_{ZF,w/o EP}$ for $\varepsilon=0.3$. Since the trapped particle portion is larger than that with $\varepsilon=0.15$, the enhancement of ZF residual level by the deeply trapped particles is more significant. 
		In one word, $R_{ZF}$ {\revise is enhanced significantly if barely passing, barely trapped or/and deeply trapped EPs are dominant}. For $\Delta\lambda=0.15$, $\varepsilon=0.15$, the barely passing EPs ($\lambda\approx0.78$) can enhance $R_{ZF}$ by a factor of $1.4\sim1.5$. As the EP becomes isotropic (large $\Delta\lambda$), the deviation of $R_{ZF}$ away from the R-H result is small as indicated by $R_{ZF}/R_{ZF,w/o EP}$ with $\Delta\lambda>0.5$. 
		
		The dependence of the ZF residual level on EP density and temperature effects are studied. The ZF residual level normalized to that w/o EPs is calculated for different values of $(n_f/n_e, T_f/T_i)$, as shown in Fig. \ref{fig:cmp_nf_Tf}. We take $\lambda_0=0.8$ and $\Delta\lambda=0.2$ where the EP leads to the enhancement of $R_{ZF}$ according to Fig. \ref{fig:ZF2d_lambda0_lambdawid}. In the long wavelength limit, $R_{ZF}/R_{ZF,w/o EP}$ is mainly affected by the EP density but the EP temperature's effect is negligible, as shown in the left frame of Fig. \ref{fig:cmp_nf_Tf}. From Eqs. \ref{eq:pola_dens0}--\ref{eq:pola_cl_dens0}, in the long wavelength limit, $K(E,\lambda,Q)\propto T_0$ and $(J_0^2-1) \propto T_0$ and thus, considering the factor $1/T_0$, it can be found  that $R_{ZF}$ is independent on $T_0$. In the moderate wavelength ($k_r\rho_{Ti}=0.1$) range, as $T_f$ increases, the classical and neoclassical polarization shielding becomes less effective and the EP contribution to $R_{ZF}$ becomes smaller as $T_f$ increases, as shown in the right frame of Fig. \ref{fig:cmp_nf_Tf}. Note that in experiments, $(n_f,T_f)$ is constraint by the accessible EP pressure. Thus in order to achieve higher $R_{ZF}$ by manipulating EPs, higher $n_f$ is preferable for both moderate and long wavelength ZFs for the same $(\lambda_0,\Delta\lambda)$ value. 
	}
	
	\subsection{ZF residual calculation using experimental EP distribution}
	\label{subsec:experiment_EP}
\begin{figure*}\centering
		\includegraphics[width=0.98\textwidth]{\myfigure/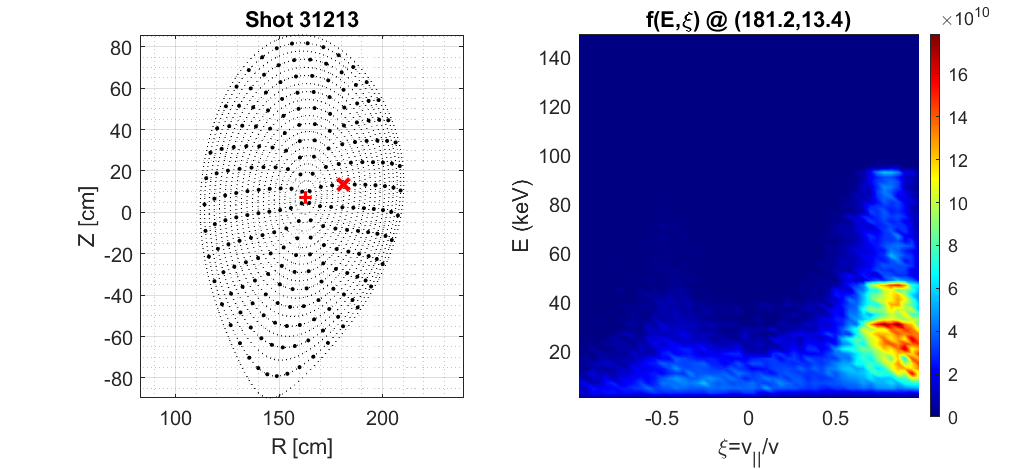}
		\caption{Left: The AUG equilibrium and the position of the EP data (red cross `$\times$'). Right: the EP distribution calculated by TRANSP-NUBEAM.}\label{fig:geometry_fExi}
	\end{figure*}

\begin{figure*}\centering
		\includegraphics[width=0.98\textwidth]{\myfigure/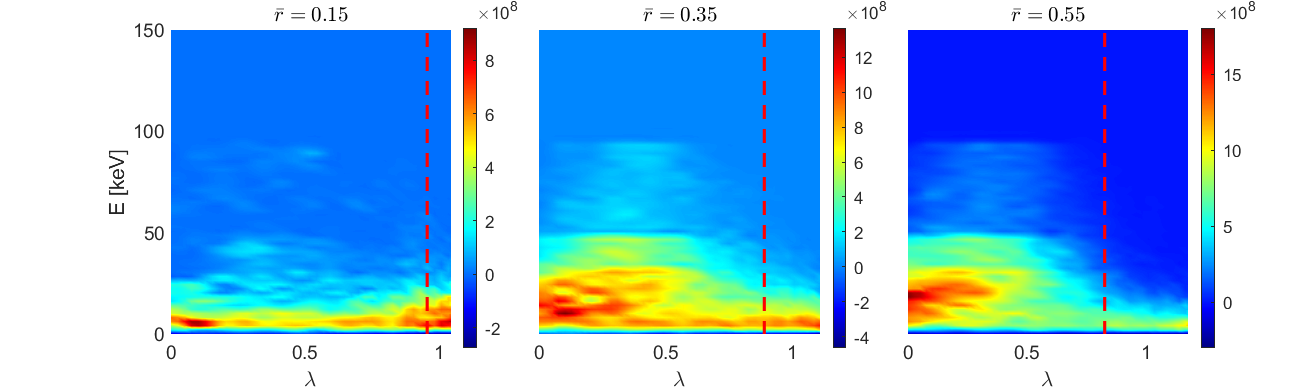}
		\includegraphics[width=0.98\textwidth]{\myfigure/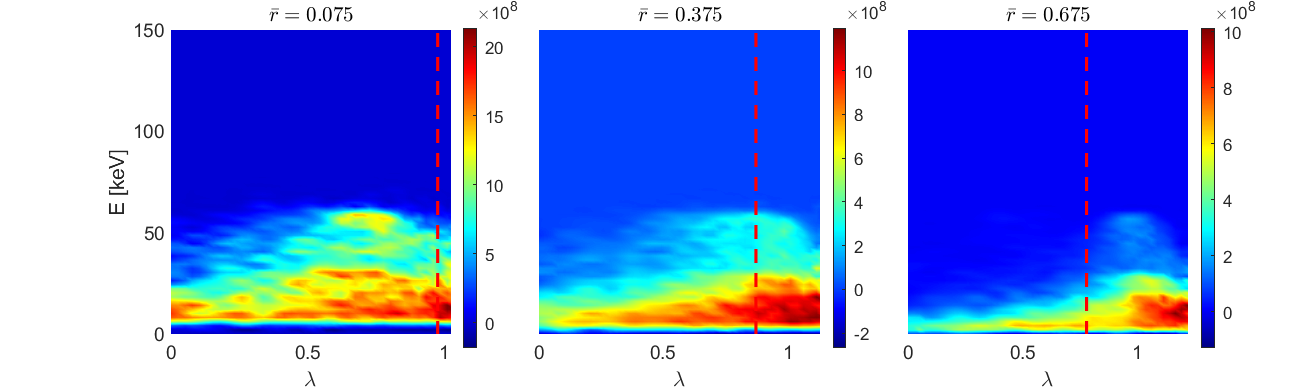}
		\caption{The EP distribution function $f(E,\lambda)$ of shot 31213 (upper) and shot 33856 (lower) at different radial locations. The red dashed line indicates the passing-trapped boundary.}\label{fig:f_E_r}
	\end{figure*}

	The ZF residual level is calculated using realistic ASDEX Upgrade EP distribution functions induced by NBI, calculated by the TRANSP-NUBEAM \cite{hawryluk1981empirical,pankin2004tokamak} code.
	The EP distribution function data is in $(R,Z,E,\xi)$ space where $(R,Z)$ corresponds to a series of positions aligned along magnetic flux coordinates as shown in Fig. \ref{fig:geometry_fExi}. 
	The EP distribution at one given point (red cross `$\times$' on the left) is shown in the right frame of Fig. \ref{fig:geometry_fExi}.
	The EP distribution function calculated  by TRANSP is in $(E,\xi)$ space and is converted to $(E,\lambda)$ space as the input of the calculation in this work. 
	{\revise
		In this work, two typical AUG cases (shot 31213 at 0.84 seconds and shot 33856 at 2.14 seconds) are chosen. For shot 33856@2.14$s$, the NBI source Q2 and short Q3 blip are applied with full injection energy of 60 keV and the injected EPs are mainly intermediate-perpendicular. 
		For shot 31213@0.84$s$, the NBI source Q7 is applied with full injection energy of 93 keV and the injected EPs are more tangential. 
		More information of the NBI source Q1-Q8 can be found in previous work \cite{geiger2013fast}. ICRH has not been applied for both cases (but our ZF residual level solver can take the EP distribution with ICRH applied as the input). 
		As shown in Fig. \ref{fig:f_E_r}, for shot 31213@0.84$s$ (left), well passing EPs are dominant near the middle minor radius, since the local maxima of the iso-contour is close to $\lambda=0$, as shown in the middle ($\bar{r}=0.35$) and right ($\bar{r}=0.55$)  frames on the top, where $\bar{r}_{tor}=\sqrt{(\psi_t-\psi_{t,axis})(\psi_{t,edge}-\psi_{t,axis})}$, $\psi_t$ is the toroidal magnetic flux function. Near the axis, some barely passing EPs and trapped EPs are also generated due to the EP drift/diffusion from the source location. 
		For shot  33856@2.14$s$ (right), barely passing EPs and trapped EPs are dominant as shown in Fig. \ref{fig:f_E_r} at the bottom. In the outer region ($\bar{r}=0.675$), the dominant EPs are trapped particles since $\lambda$ at the local maxima of the iso-contour is larger than the passing-trapped boundary (the red dashed line).
	}
	
\begin{figure*}\centering
		\includegraphics[width=0.8\textwidth]{\myfigure/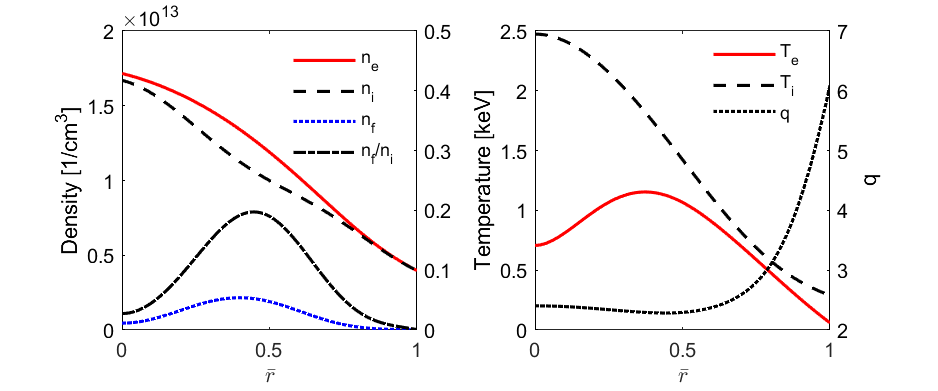}
		\includegraphics[width=0.8\textwidth]{\myfigure/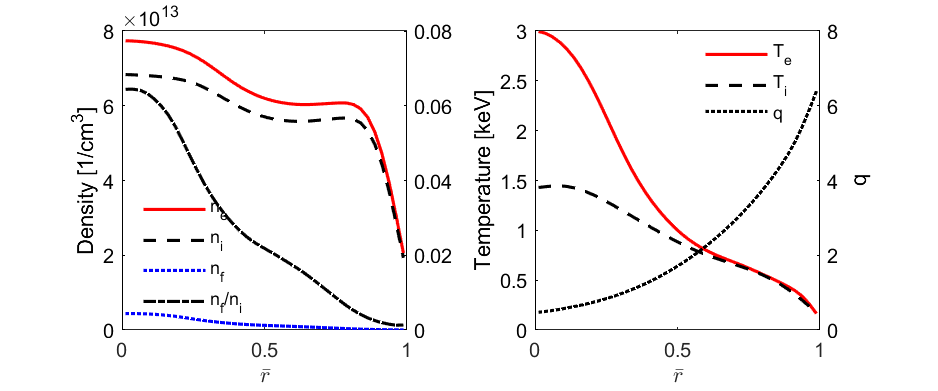}
		\caption{The density, temperature, safety factor profiles of the AUG shot 31213 (upper) and shot 33856 (lower). For shot 31213, $n_i$ is constructed from $n_i=n_e-n_f$. Profiles of shot 33856 are all from TRANSP data. Besides the $q$ profile, another input profile ($n_f/n_i$) for calculation of the ZF residual level is also shown with Dash-dot line in the left frame. }\label{fig:ntqprofiles}
	\end{figure*}

\begin{figure*}\centering
		\includegraphics[width=0.48\textwidth]{\myfigure/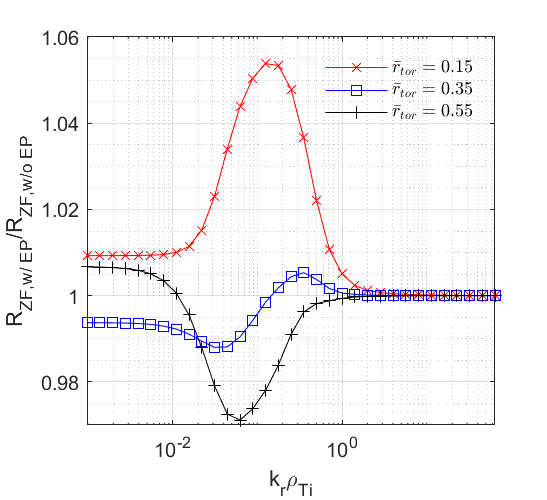}
		\includegraphics[width=0.48\textwidth]{\myfigure/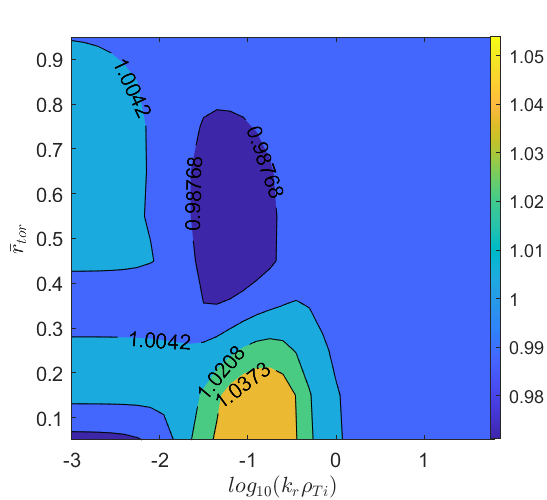}
		\includegraphics[width=0.48\textwidth]{\myfigure/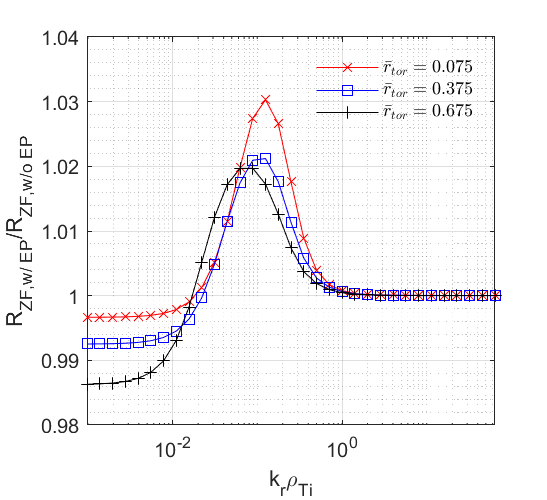}
		\includegraphics[width=0.48\textwidth]{\myfigure/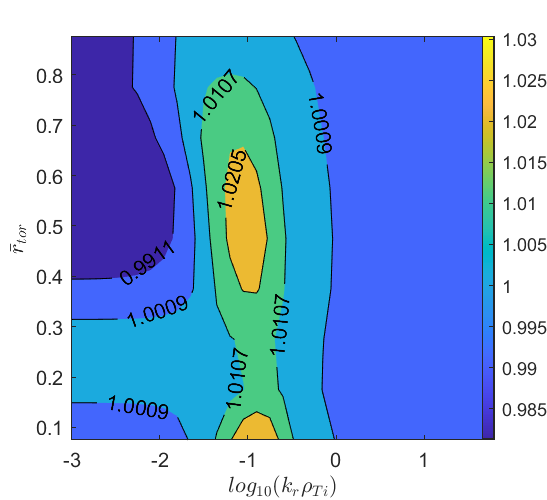}
		\caption{The ZF residual level for shot 31213 (top)) and shot 33856 (bottom)) with different values of radial location and $k_r\rho_{Ti}$. }\label{fig:zfres_EP2d3d}
	\end{figure*}

	{\revise
		The numerical solver has been developed recently to treat arbitrary experimental EP distributions when solving the ZF residual level with the input parameters $(\varepsilon, q, k_r\rho_{Ti}, T_f/T_i, n_f/n_e)$. Numerical interfaces in the ZF residual level solver have been implemented to collect the parameters from the TRANSP and TRANSP-NUBEAM data. The EP distribution function data $f(E,\lambda,r)$ is reconstructed from the TRANSP-NUBEAM data using spline functions. In our calculation, concentric circular magnetic surfaces are assumed with the inverse aspect ratio calculated according to $\varepsilon=(R_{max}-R_{min})/(R_{max}+R_{min})$ where $R_{max}$ and $R_{min}$ are the maximum and minimum major radii in a given magnetic surface.
		The reference EP and thermal ion density and temperature as well as the safety factor are chosen according to the experimental values as shown in Fig. \ref{fig:ntqprofiles}. 
	}
	The enhancement factor of ZF residual level $R_{ZF,w/\;EP}/R_{ZF,w/o\;EP}$ is calculated and is shown in Fig. \ref{fig:zfres_EP2d3d}. 
	{\revise
		For shot  31213, except in the near-axis region, well passing EPs are dominant as shown in Fig. \ref{fig:f_E_r} and the main effects on ZF residual is suppressing, especially in the intermediate wavelength range ($k_r\rho_{Ti}\sim 0.1$). The ZF residual at $\bar{r}=0.55$ is suppressed by $\sim3\%$. In the near-axis region where significant barely passing and trapped EPs are present ($\lambda$ close to $1$), the ZF residual level is enhanced by $\sim5.5\%$ as indicated by the red line, at $k_r\rho_{Ti}\sim0.1$. The ZF residual enhancement $R_{ZF,w/ EP}/R_{ZF,w/o EP}$ in $(\bar{r},k_r\rho_{Ti})$ space is shown in the right frame on the top. In the $0.35<\bar{r}<0.75$ region, EPs suppress the ZF residual level by $1\sim2\%$. EPs' effects near the edge ($\bar{r}>0.8$) is small since $n_{f}/n_i$ is small ($n_f/n_i<0.03$) as indicated in the left top frame of Fig. \ref{fig:ntqprofiles}.
		For shot  33856, the main EP effects on ZF residual is enhancement in the intermediate wavelength range $k_r\rho_{Ti}\sim0.1$ and suppression in the long wavelength range $k_r\rho_{Ti}<10^{-2}$, in most of the radial locations. ZFs residual level can be enhanced by $\sim2\%$ at $\bar{r}=0.375$ for $k_r\rho_{Ti}\sim0.1$. As shown in the lower frame of Fig. \ref{fig:f_E_r}, significant portion of EPs are barely passing, barely trapped and/or deeply trapped. In contrast to shot  31213,  the well passing EPs are few in shot  33856. The absence of the well passing EPs is correlated with the enhancement of ZF residual, which is consistent with the results of shot  31213 and the the studies using the model EP distributions in Section \ref{subsec:model_EP}. The intermediate trapped EPs ($\lambda\sim1$) also take a significant portion for shot  33856, which is expected to suppress the ZF residual level found in Section \ref{subsec:model_EP}. The results of the ZF residual level suggest that the overall effect is mainly dominated by barely passing/trapped and deeply trapped EPs compared to the intermediate EPs. 
	}
	
	{\revise
		The analysis using experimental EP distribution also sheds some light on improving the ZF residual enhancement by manipulating the EP source and plasma parameters. Note that for shot  33856, the EP portion is low ($n_f/n_i\sim2.96\%$ at $\bar{r}=0.375$) due to the higher ion density compared with the low ion density shot 31213. For example, by increasing the ratio $n_f/n_i$ of shot  33856 four times (by either decreasing the ion density or increasing EP density in experiments), the ZF residual is expected to be enhanced by $9.1\%$ at $\bar{r}_{tor}=0.375$ and by $13.7\%$ at $\bar{r}_{tor}=0.025$ from our calculation. 
		Higher values of EP concentration (e.g. $n_f/n_i\sim0.4$) have been obtained in other ASDEX Upgrade discharges such as shot 30809 as reported in previous work \cite{weiland2017phase} and the systematic analysis using various experimental parameters will be our future work.
		In addition, by optimizing the NBI injection parameters such as selecting the NBI with different injection angles and injection energy, the anisotropy of EPs can be adjusted in order to reduce the portion of well passing or intermediate trapped EPs, which is preferable for the enhancement of the ZF residual level. 
		Nevertheless, using the present AUG cases, for the intermediate wavelength $k_r\rho_{Ti}\approx 10^{-1}$, the EP effects on ZF residual can be suppressed or enhanced by $-3\%\sim+5.5\%$. 
	}
	In a nonlinear system where the ZF is driven via nonlinear interaction, the corresponding ZF residual level changes by the same factor if the nonlinear drive stays at the constant level. 
	{\revise
		Nonlinear simulations demonstrate that ZFs  can be generated by Alfv\'en eigenmode (AE) via wave-wave nonlinearity ( \cite{todo2010nonlinear,chen2018zonal} and references therein). In turn, ZFs can reduce the AE saturation and the EP transport level. The wave-particle nonlinear interaction and the related power exchange as well as particle transport in phase space can modify the distribution of the thermal ions and EPs and thus change the ZF residual level. For more realistic evaluation of the ZF residual level, comprehensive consideration of the nonlinear EP-AE-ZF system  needs to be considered in future. 
	}
	Note that for cases with lower EP pitch scattering due to, e.g., higher EP  energy, the distribution width in $\lambda$ direction can be smaller in other devices \cite{kamada2011plasma,bierwage2017mhd}, and the EP effects on ZF residual can be more significant for the same portion of EPs. 
	
	\section{Summary and conclusions}
	\label{sec:conclusion}
	In this work, the effects of anisotropic EPs on ZF residual level are studied. The equations for classical and neoclassical polarization densities are derived and implemented for arbitrary distribution functions.  
	{\revise
		Model EP distributions or experimental ones are adopted to demonstrate the underlying physics of the ZF enhancement or suppression and to evaluate the possible impact on experiments.
	} 
	The main results are summarized as follows.
	\begin{enumerate}
		\item As is well known \cite{rosenbluth1998poloidal}, for Maxwellian particles and in the long wavelength limit, the neoclassical polarization density is contributed by passing and trapped particles, with similar magnitude, of order $(k_r\rho_{Ti}q)^2/2/\varepsilon^2$ but with opposite signs; and with the passing and trapped particle contributions cancel partly, one can get $\delta n^{pol}_{nc}\approx1.6n_0e\delta\phi(k_r\rho_{Ti}q)^2/(2\varepsilon^{1/2}T_0)$ . As shown in this work, for anisotropic EPs, the cancellation of trapped and passing particle contribution is different and the polarization density is different, which indicates different levels of residual ZFs.
		\item {\revise The anisotropic EPs have more significant impact on intermediate ($1<k_r\rho_{Ti}<0.1$) and long wavelength ($k_r\rho_{Ti}<1$) ZFs. For short wavelength ($k_r\rho_{Ti}>1$) ZFs, EPs' effects are not significant.}
		\item In the long wavelength ($k_r\rho_{Ti}<1$) range, passing EPs (except barely {\revise passing} EPs) and intermediate trapped (neither barely trapped nor deeply trapped) EPs lead to the decrease of ZF residual level. Barely passing, barely trapped or deeply trapped EPs lead to increase of ZF residual level.
		\item {\revise For ASDEX Upgrade plasmas, EP distribution functions in two selected shots can bring in $-3\%\sim+5.5\%$ mitigation/enhancement in ZF residual level, depending on the EP distribution functions. Higher ZF enhancement (e.g., $\sim10\%$) is expected by appropriate manipulation of EP and equilibrium parameters, such as by raising EP density and increasing barely passing/trapped or deeply trapped EPs. }
	\end{enumerate}
	Future work relies on efforts in modeling, simulations and experiment sides. More studies include, but are not limited to,
	\begin{enumerate}
		\item the theoretical analysis and numerical modeling with the consideration of more realistic tokamak geometry/EP source \cite{bilato2011simulations,weiland2018rabbit} and more self consistent nonlinear drive of ZFs \cite{chen2000excitation};
		\item the simulations of ZFs with non-Maxwellian EP distributions using gyrokinetic codes \cite{di2018non,lanti2020orb5,lu2019development,lu2021development} or hybrid MHD-gyrokinetic codes \cite{todo2010nonlinear}, with the background turbulence or EP driven modes self consistently simulated as the primary instabilities;
		\item the experimental studies of EP effects on the ZF level and the turbulence control, by manipulating the EP parameters, especially the EP pitch and the EP energy, using NBI or ICRH.
	\end{enumerate}
	Noticing the theoretical and numerical analyses in this work, the effect of the anisotropic EPs on ZF residual level can be important for tokamak plasmas for which the EP distributions are narrow in pitch \cite{kamada2011plasma,pinches2018progress,bierwage2017mhd}, as to be studied in the future. 
	
	
	
	\appendix
	\section{Gyrokinetic solution of the axisymmetric perturbation ($n=0$ for arbitrary distribution function)}\label{app:gkf1derivation}
	{\revise The gyrokinetic solution for Eq. \ref{eq:gk0} is solved perturbatively using the small parameter $|\omega/\omega_{tr}|\ll1$. 
		While early work is for Maxwellian distribution \cite{rosenbluth1998poloidal,xiao2006short}, in this work, we consider arbitrary distribution functions.
		For axisymmetric perturbations, $\omega_d=v_\parallel\partial_{\parallel}Q$,   $Q=IS'v_\parallel/\omega_{c}$, $I=B_\phi R$, $S=S(\psi)$, $S'=dS/d\psi$, and $\psi$ is the poloidal flux function.
	} 
	With the expansion of Eq. \ref{eq:gk0} in $\omega/\omega_{tr}$, the zeroth order equation is
	\begin{eqnarray}
	\left[v_{||}\partial_{||}+i\omega_d\right] \delta K_0 
	=0\;\;,
	\end{eqnarray}
	whose solution is $\delta K_0=\delta H\exp(-iQ)$, where $\partial_\parallel\delta H=0$.
	The first order equation is 
	\begin{eqnarray}\label{eq:deltaK0}
	\left[v_{||}\partial_{||}+i\omega_d\right] \delta K_1 
	= -\partial_t\delta K_0+\frac{e_s}{T_{0s}} \kappa F_{0s}  J_0(a_s)\partial_t\delta\phi\;\;.
	\end{eqnarray}
	After bounce average operation $\bar{A}=\oint (dl/v_\parallel)A/\oint (dl/v_\parallel)$, where $l$ is the coordinate along $\bf b$, Eq. \ref{eq:deltaK0} leads to the solution to $O(\omega/\omega_{tr})$,
	\begin{eqnarray}
	\delta H=\frac{e}{T_0}\kappa F_{0} \overline{J_0 e^{iQ}\delta\phi}\;\;,\,\;\;
	\delta f=\frac{e\delta\phi}{T_0}\kappa F_0(J_0 e^{-iQ}\overline{J_0 e^{iQ}}-1)\;\;.
	\end{eqnarray}
	{\revise 
		$F_0$ is an arbitrary distribution function and $\kappa F_{0s}\equiv -(T_{0s}/m_s)\partial F_{0s}({\bf R},E,\mu)/\partial{E}$. In the $(E,\lambda)$ coordinates, where the pitch is defined as $\lambda=v^2_\perp B_0/(v^2B)$, $\kappa F_0$ is obtained noticing
		\begin{eqnarray}
		\label{eq:dfdE2dfdlambda}
		\left.\frac{\partial}{\partial E}\right|_\mu F_0 = \left.\frac{\partial}{\partial E}\right|_\lambda F_0 - \left.\frac{\lambda}{E}\frac{\partial}{\partial\lambda}\right|_E F_0\;\;.
		\end{eqnarray} 
	}
	
	\section{Reduction to Rosenbluth-Hinton results in long wavelength limit for Maxwellian distribution}
	\label{app:compare_RH}
	For the comparison of Eqs. \ref{eq:npol0} and \ref{eq:nclpol0} with previous work, we calculate the classical/neoclassical polarization $\varepsilon_{cl/nc}^{pol}$
	\begin{eqnarray}\label{eq:define_polarization}
	\varepsilon^{pol}_{cl/nc}\langle k_r^2\rangle\delta\phi\equiv-4\pi e_s\langle\delta n_{cl/nc}^{pol}\rangle\;\;.
	\end{eqnarray}
	Then Eqs. \ref{eq:npol0} and \ref{eq:nclpol0} yield,
	\begin{eqnarray}\label{eq:polacl1}
	\varepsilon_{cl}^{pol}=\frac{\omega_{ps}^2}{\omega_{c0s}^2}\frac{\pi}{2} v_{Ts}^3\sum_{\sigma}\int d\lambda\oint\frac{d\theta h}{2\pi}\int_0^\infty d\bar{E}\kappa\bar{F}\bar{{E}}^{3/2}
	\left(	\frac{1}{\xi}	-\xi	\right)
	\;\;,\\
	\label{eq:polanc1}
	\varepsilon_{nc}^{pol}=\frac{\omega_{ps}^2}{\omega_{c0s}^2}\frac{q^2}{\varepsilon^2}\pi v_{Ts}^3\sum_{\sigma}\int d\lambda\oint\frac{d\theta}{2\pi}\int_0^\infty d\bar{E}\kappa\bar{F}\bar{{E}}^{3/2}
	\left(
	h^2\xi-\frac{2\pi h}{\oint d\theta/\xi}
	\right)\;\;.
	\end{eqnarray}
	
	Equations \ref{eq:polacl1} and \ref{eq:polanc1} are consistent with  the previous results for Maxwellian distribution ($\kappa=1$) \cite{rosenbluth1998poloidal,xiao2006short}, i.e.,
	\begin{eqnarray}
	\label{eq:varepscl0}
	&&\left\langle\varepsilon_{cl}^{pol}\right\rangle=\frac{\omega_{ps}^2}{\omega_{c0s}^2}\left(1+\frac{3}{2}\varepsilon^2\right)\;\;,
	\\
	\label{eq:vareps0}
	&&\left\langle\varepsilon_{nc}^{pol}\right\rangle=\frac{\omega_{ps}^2}{\omega_{c0s}^2}\frac{q^2}{\varepsilon^2}I_{tot}\;\;,\;\;
	I_{tot}=\sum_{p,t}[I_2-\sigma_pI_1] \\
	\label{eq:T12}
	&&I_{1,2}\equiv\frac{3}{2}\int d\lambda T_{1,2}\;\;,\;\;
	T_1\equiv\frac{2\pi}{\oint d\theta/\xi};\;,\;\;
	T_2\equiv\frac{\oint d\theta\xi}{2\pi}\;\;,
	\;\;,
	\end{eqnarray}
	where $\omega_{ps}=4\pi ne_s^2/m_s$, $\oint d\theta$ indicates the integral along the unperturbed orbit, $\sigma_p=0, 1$ for trapped and passing particles respectively, the factor $(3/2)\varepsilon^2$ is due to the flux surface average. In small $\varepsilon$ limit, Eq. \ref{eq:vareps0} yields the well known Rosenbluth-Hinton result 
	$	\varepsilon_{nc}^{pol}=
	1.6({\omega_{ps}^2}/{\omega_{c0s}^2})({q^2}/{\varepsilon^{1/2})}
	$ \cite{rosenbluth1998poloidal} noticing $I_1^p\approx1-1.6\varepsilon^{3/2}$, $I_2^p\approx1-1.2\varepsilon^{3/2}$, $I_2^t\approx1.2\varepsilon^{3/2}$ \cite{xiao2007collisional}.

	\section*{Acknowledgments}
	Discussion with and input from Lu Wang, T.S. Hahm and R. Bilato are acknowledged by Z. X. Lu. 
	This work is supported by the EUROfusion Enabling Research Projects WP19-ER/ENEA-05 (MET) and WP17-ER/MPG-01 (NLED). This work has been carried out within the framework of the EUROfusion Consortium and has received funding from the Euratom research and training programme 2014-2018 and 2019-2020 under grant agreement No 633053. The views and opinions expressed herein do not necessarily reflect those of the European Commission.
	\\
	
	
	
	\providecommand{\newblock}{}

	
	
\end{document}